\documentclass[a4paper,twoside]{article}
\newcommand{\affil}[1]{$^{\rm #1}$}
\usepackage[authoryear]{natbib}
\bibpunct{(}{)}{;}{a}{}{,}
\usepackage{graphicx}
\usepackage{epstopdf}
\usepackage{caption}
\usepackage{url}  
\usepackage{amsmath}
\usepackage{amssymb}
\date{} %Please leave the date blank
\date{
{\small \it PASA received 2010 Mar 13; accepted 2010 June 9}
}
\title{\large\bf\flushleft The Zadko Telescope: A Southern Hemisphere
 Telescope for Optical Transient Searches, Multi-Messenger Astronomy
 and Education}
\author{\parbox{\textwidth}{\flushleft
\vspace{-0.5cm}
{\it D.~M. Coward\affil{A,E}, M.~Todd\affil{B},
 T.~P. Vaalsta\affil{A},  M. Laas-Bourez\affil{D,A},
 A. Klotz\affil{C,D},\\
 A. Imerito\affil{A}, L. Yan\affil{A}, P. Luckas\affil{A},
 A.~B. Fletcher\affil{A}, M.~G. Zadnik\affil{B},\\
 R.~R. Burman\affil{A}, D.~G. Blair\affil{A}, J. Zadko\affil{A},
 M. Bo\"{e}r\affil{D}, P. Thierry\affil{C},\\
 E.~J. Howell\affil{A}, S. Gordon\affil{A}, A. Ahmat\affil{A},
 J.~A. Moore\affil{A}, and K. Frost\affil{A}
}\\
\vspace{4.0mm}
{\small
 \affil{A}\,School of Physics, M013,
 The University of Western Australia, 35 Stirling Hwy,\\
 \hspace{1.8mm} Crawley, WA 6009, Australia}\\
\vspace{0.5mm}
{\small \affil{B}\,Department of Imaging and Applied Physics,
 Bldg 301, Curtin University of Technology,\\
 \hspace{1.8mm} Kent St, Bentley, WA 6102, Australia}\\
\vspace{0.5mm}
{\small \affil{C}\,CESR, Observatoire Midi-Pyr\'{e}n\'{e}es, CNRS,
  Universit\'{e} de Toulouse,\\
 \hspace{1.8mm} BP 44346, F--31028 Toulouse Cedex 4, France}\\
{\small \affil{D}\,Observatoire de Haute-Provence,
  F--04870 Saint Michel l'Observatoire, France}\\
{\small \affil{E}\,Corresponding author. E-mail: coward@physics.uwa.edu.au} }
}
\begin{document}
\twocolumn[
\begin{changemargin}{.8cm}{.5cm}
\begin{minipage}{.9\textwidth}
\vspace{-1cm}
\maketitle
%
%
%%%%%%%%%%%%%     ABSTRACT    %%%%%%%%%%%%%
%Abstract of no more than 200 words here.
% ABF (21 Feb 2010): <= 200 words.
\small{ {\bf Abstract:} The new 1-m f/4 fast-slew Zadko Telescope
 was installed in June 2008 about 70 km north of Perth, Western Australia.
 It is the only metre-class optical facility at this southern latitude
 between the east coast of Australia and South Africa, and
 can rapidly image optical transients at a longitude
 not monitored by other similar facilities.
 We report on first imaging tests of a pilot program of
 minor planet searches, and Target of Opportunity observations
 triggered by the {\it Swift} satellite.
 In 12 months, 6 gamma-ray burst afterglows were detected,
 with estimated magnitudes;
 two of them, GRB 090205 ($z = 4.65$) and GRB 090516 ($z = 4.11$),
 are among the most distant optical transients imaged by
 an Australian telescope.
 Many asteroids were observed in a systematic 3-month search.
% at a rate of 0.007 deg$^{-2}$ hr$^{-1}$ of observing.
 In September 2009, an automatic telescope control system
 was installed, which will be used to link the facility to
 a global robotic telescope network;
 future targets will include
 fast optical transients triggered by high-energy satellites,
 radio transient detections, and
 LIGO gravitational wave candidate events.
 We also outline the importance of the facility as
 a potential tool for education, training, and public outreach.
}

%%%%%%%%%%%%%     KEYWORDS    %%%%%%%%%%%%%
\medskip{\bf Keywords:} instrumentation: miscellaneous ---
 gamma rays: bursts --- minor planets, asteroids --- telescopes
% Please write all keywords in lower case. PASA uses the
% standard list of subject headings adopted by
% The Astrophysical Journal and available from
% http://www.journals.uchicago.edu/ApJ/keywords_text.html.
% Keywords are separated by em-dashes, i.e. ---

%%%%%%%%DO NOT EDIT%%%%%%%%%%%%
\medskip
\medskip
\end{minipage}
\end{changemargin}
]
\small
%%%%%%%%EDIT FROM HERE%%%%%%%%%%%%

%=======================================================================
%
\section{Introduction}
\label{sectintro}
% Please see the PASA Style Guide for help with correct layout
% for your manuscript.
% Examples of tables and figures are given below.

The Universe is teeming with fleeting transients, visible across the
entire electromagnetic spectrum. In the optical, some last for only
a few seconds, while others vary in brightness over years.
%
% The richness of the transient sky is demonstrated by a recent report
% from the the University of Arizona Catalina Sky Survey (CSS)
% \footnote{\url{http://uanews.org/node/26922}}.
% A surprising number of optical transients was discovered over
% a 17 month period: more than the team can characterise.
%   The team also found about 100 other highly variable optical
% sources, including active galactic nuclei, high proper motion stars
% and sources that remain unknown. Given that the CSS primary driver
% is the search for Near Earth Objects (NEOs), these other
% serendipitous discoveries are even more remarkable and are
% providing new insight into the so-called `transient Universe'.
%
Sources classed as transient include
blazars, cataclysmic variable stars, stellar flares,
supernovae, and hypernovae;
many of the recently discovered sources are not easily explained. 
 
Some transients, such as gamma-ray bursts (GRBs), have complex
emissions spanning the gamma, X-ray and optical bands, and are
possibly also strong gravitational wave and neutrino sources.
In `multi-messenger' astronomy (see Sathyaprakash \& Schutz 2009),
coordinated observations of the different kinds of radiation and
information carriers emitted by such sources offers the opportunity
to probe the physics of these exotic transients, by studying their
diverse emissions over a range of energies and timescales,
from seconds to days.
%
% Another superb facility engaged in exploring the optical transient
% sky is The Palomar Transient Factory (PTF). It is a fully-automated,
% wide-field survey using a new 8.1 square degree camera installed on
% a 48-inch telescope at Palomar Observatory \citep{Law09},
% while colors and light curves for detected transients are obtained
% with the automated Palomar 60-inch telescope.
% The PTF operation strategy is designed to probe existing gaps in
% the transient phase space \citep{Rau09}, and to search for
% theoretically predicted phenomena, such as fallback supernovae,
% macronovae, and the afterglows of `off-axis' gamma-ray bursts.
% During commissioning in 2009, several tens of transients were
% discovered over a period of days. 
%
% The PTF has been described as a pathfinder to the ambitious
% 8.4 m Large Synoptic Survey Telescope
% \footnote{\url{http://www.lsst.org/lsst}}
% (LSST): it will survey the entire sky in multiple colors every week
% using a three-billion pixel CCD camera.
% Another wide field instrument Pan-STARRS
% \footnote{\url{http://pan-starrs.ifa.hawaii.edu/public/home.html}}
% (Panoramic Survey Telescope And Rapid Response System) is a survey
% instrument (now operational) that will conduct astrometry and photometry
% of much of the sky on a continuous basis using a gigapixel camera with
% FOV of $3^{\mathrm o}$.
% In Australia, the SkyMapper Telescope \citep{Keller07}, just commencing
% first observations, is a 1.3-metre multi-colour survey telescope
% employing a 16k $\times$ 16k CCD mosaic with $0.5^{\mathrm '}$ pixels
% covers 5.7 square degrees.
%

There are new and upcoming facilities which are expected to discover
a huge number of transient objects -- these include
large optical and radio facilities, space-based detectors, and
the next generation of gravitational wave and neutrino observatories.
The 1-m Zadko Telescope, with its 1$^{\circ}$ field of view,
is perfectly poised to study the transient sky, by following up
and monitoring the many transients that will be detected by these
facilities in the coming decade. \\
%
% The global distribution of large (8-10 metre class) telescopes is dictated
% by geography and optimal site conditions, leading to the clustering of
% superb facilities.  For the case of 0.5- to 1-m class telescopes, the
% argument for locating the facilities at often remote locations is not
% as compelling unless existing infrastructure can be used, as is the case
% for La Silla Paranal Observatory in Chile, operated by ESO. 
%
% Importantly, metre class-telescopes are finding a niche role where it is
% advantageous to have a broad geographical distribution.
% Larger facilities can miss optical transients that fade rapidly, due to
% the wide separation between these facilities.
%
\indent
The study of transient optical sources that vary on timescales of
seconds to hours is an exciting and growing field, boosted by
the widespread use of small to medium,
rapid-response automated telescopes.
Many of the smaller instruments are being networked to coordinate
observations to allow near-continuous monitoring of
rapidly fading (or flaring) optical transients. 

\begin{figure}[t!]
%\setcaptionwidth{3in}
% \includegraphics[scale=0.29]{telescope_v2.pdf}
\includegraphics[scale=0.29]{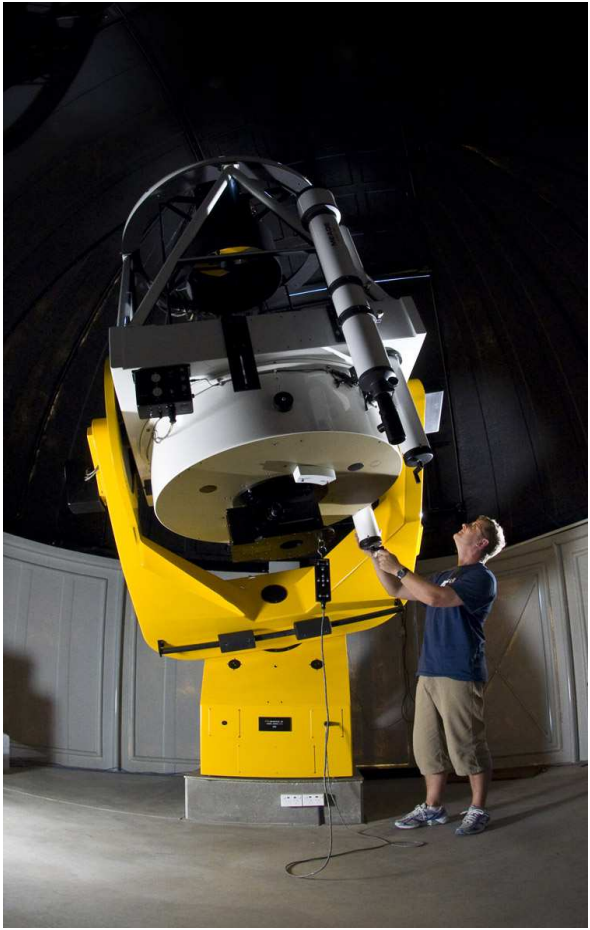}
\caption{ \small
Zadko Telescope: a new 1-m f/4 Cassegrain reflector,
located 67 km north of Perth.
Built by DFM Engineering, Inc.,
and donated to UWA by resource company Claire Energy,
it was opened by Australia's Chief Scientist on 2009 April 1. }
\label{figdome}
\end{figure}

The Zadko Telescope (ZT), shown in Figure \ref{figdome},
is just starting first operations as a networked robotic observatory.
Made possible by a philanthropic donation to
the University of Western Australia (UWA),
it is a new resource for research, training, and science education.
It is colocated with a science and astronomy outreach facility,
and with the Australian International Gravitational Observatory (AIGO). 

The leaders of the French robotic telescope network
TAROT\footnote{\url{http://tarot.obs-hp.fr}}
(T\'{e}lescopes \`{a} Action Rapide pour les Objets Transitoires)
\citep*{Klotz08a}
recognised that the ZT is uniquely located in a region devoid
of similar facilities. This led to an ongoing partnership
to robotise the facility for optical transient science,
using the TAROT robotic observatory control system.
The joint science capability of the ZT will be substantially
extended by this partnership in two ways:
firstly, the ZT will provide deeper observations,
so it can follow up on transients fainter than TAROT can detect;
secondly, it will increase the overall detection rate, due to
its wide longitudinal separation from the TAROT Chile site. 

The ZT, in partnership with other facilities,
such as TAROT and SkyMapper,
can make valuable contributions to
frontier multi-messenger astronomy,
including searches for optical counterparts of
radio, X-ray, and
LIGO gravitational wave (GW) transients.

The LIGO\footnote{\url{http://www.ligo.caltech.edu}}
and VIRGO\footnote{\url{http://www.virgo.infn.it}}
interferometers are now entering a sensitivity regime
where GW emissions from coalescing compact binaries
should be detectable out to 10s of Mpc;
the next-generation Advanced GW detectors will expand
this horizon out to 100s of Mpc \citep{Abadie10}.
This presents an enormous opportunity to combine
optical transient searches with GW burst searches.
From early 2010, the ZT is being employed by LIGO as
a test facility for optical follow-up of large error boxes
($\thickapprox$10 deg$^{2}$). The first tests and methods of this
pilot program will be reported in a later publication.

The ZT has already allowed scientists to engage with the broader community
on several research proj-ects that have captured the public imagination.
Furthermore, the ZT has opened up new research opportunities that have
the potential to contribute to Australia's already important role
in optical transient astronomy. In this paper, we describe the status of
the facility, and its planned role in several current and emerging transient
astronomy science programs.
%
% The facility will also be used for solar system astronomy:
% NEO searches are gaining a higher priority with NASA, and
% the ZT can potentially play an important role here.
% A Zadko pilot NEO search commenced in early 2009, and
% we present preliminary results from the first few months of
% this search in Section \ref{sectminplanets}.
%

%=======================================================================
%
\section{The Zadko Telescope}
\label{sectzadkotel}
The ZT\footnote{\url{http://www.zt.science.uwa.edu.au}}
is a new 1-m f/4 Cassegrain telescope,
built by DFM Engineering,
Inc.\footnote{\url{http://www.dfmengineering.com}},
and is situated in the state of Western Australia
at longitude $115^{\circ} 42' 49''$ E,
latitude $31^{\circ} 21' 24''$ S,
and at an altitude of $47$\,m above sea level.
Located $67$\,km north of Perth, the site has a
typical `Californian' climate ---
hot, dry summers with mostly clear skies,
and mild, wet winters,
with a mean of 83 clear days per year.
The closest towns are each $19$\,km away, with
small populations (1628 \& 531 in the 2006 Census);
the site is not strongly affected by man-made lighting.
% the site is dark at night.
%
% to Universal Time.)  The site is about 19km west of the town of
% Gingin, 19km north east of the harbour town of Two Rocks and 67km
% north of the capital city of Perth.  
%

The ZT is an equatorial, fork-mounted telescope
with a primary mirror clear aperture of $1007.0$\,mm,
and a system focal length of $4038.6$\,mm.
Its fast optics have a low f-ratio, and
a flat and wide field of view.
It uses state-of-the-art friction drives to achieve
high stiffness and excellent tracking on both axes.
The fork mount provides for a convenient and compact telescope,
able to carry a maximum instrument load of $200$\,kg, and
requiring minimal counterweights
for fast dynamic response --- with a maximum
slew speed of $3.0^{\circ}$\,s$^{-1}$.
The observatory's Sirius dome
rotates at $3.7^{\circ}$\,s$^{-1}$ on average.
%
% \begin{figure}[th!]
% \setcaptionwidth{3in}
% \includegraphics[scale=0.29]{telescope_v2.eps}
% \includegraphics[scale=0.32]{site-345pt-width2.eps}
% \caption{ \small
% {\bf Top:} The Zadko Telescope: a DFM 1-m Cassegrain telescope,
% located about 80 km north of Perth,
% donated to UWA by resource company Claire Energy,
% and opened by Australia's Chief Scientist on 1 April 2009.
% {\bf Bottom:} The ZT observatory dome and weather monitoring instruments.}
% \label{figdomesite}
% \end{figure}
%

The primary and secondary mirrors are spaced with
Invar rods to minimise focus shift with temperature.
The carbon-fibre technology within the focus housing
provides additional thermal compensation of the optical
spacing\footnote{This focus housing delivered by DFM Engineering, Inc.,
is the first based on a new carbon-fibre design.}.
For a slowly changing temperature, the optical assembly has
an essentially zero temperature coefficient.

The ZT is mounted on a vibration-isolated concrete pier,
and housed in a 6.7-m diameter fibre-glass dome.
Mechanical vibrations transferred from the observatory floor
to the mount do not exceeded $1$\,arcsec,
and are damped within $1$\,s.
% The dome has two hydraulically-controlled shutters
% powered by battery, and recharged by a solar panel.
The dome has two shutters: the upper is driven by electric motor,
and the lower is a hydraulically-controlled drawbridge;
both are powered by battery,
which is recharged by a solar panel.
The dome is interfaced with the ZT using the
ASCOM (AStronomy Common Object Model) protocol,
and is controlled via the French Robotic Observatory Control System,
installed in 2009 September (see Section \ref{sectrocs}). 

%=======================================================================
%
\subsection{Optical Design}
\label{sectoptdesign}
The ZT optical design is a hybrid related to
the traditional Ritchey-Chr\'{e}tien (RC), but with
a small, thin, 4th-order refracting field corrector-flattener.
The two mirrors form a modified RC system, but are
significantly more aspheric --- especially the secondary mirror.
The optical design also incorporates the effects of the glass in
the filter wheel and CCD vacuum window.

In this design, incident light is reflected by a
light-weight f/2.3 primary mirror of $1007.0$\,mm diameter (clear aperture),
% lightweight f/2.3 primary mirror of 1007.0 mm diameter clear aperture,
an f/4.9 secondary mirror of $469.9$\,mm diameter,
% an f/4.9 secondary mirror of 469.9 mm diameter (clear aperture),
and passes through
the very fast and thin $6.35$\,mm-thick 4th-order
Schmidt corrector-flattener plate of $236.2$\,mm diameter,
% Schmidt corrector-flattener plate of 236.2 mm diameter (clear aperture),
placed $126.2$\,mm above the primary mirror surface.
The resulting system focal ratio is f/4.0105 at the Cassegrain focus,
which is nominally $467.9$\,mm below the primary mirror surface.

Focusing is achieved at the secondary mirror housing, which
allows $77$\,mm of back focus travel.
When properly focused, 80\% of the incident light falls
within $0.6$\,arcsec (in vacuo).
The ZT is designed to image on detectors
as wide as $99$\,mm in diameter,
giving an unvignetted $1.4^{\circ}$ field of view (FOV),
and the image spot size nominally
remains $<1$\,arcsec,
at a deviation of $0.65^{\circ}$ off-axis.
\subsection{Low-noise CCD Imager}
\label{sectandorccd}
The main ZT imaging detector is
a large-area, low-noise CCD camera,
the iKon-L with a DW436-BV back-illuminated sensor,
built by Andor Technology
PLC\footnote{\url{http://www.andor.com}}.
This thermoelectrically-cooled CCD
has $2048\times2048$\,$13.5$\,$\mu$m\,square pixels,
corresponding to $0.69$\,arcsec
at the ZT f/4.0 Cassegrain focus;
the CCD FOV is $23'.6\times23'.6$.
The pixel well-depth is $\thickapprox100,000$\,$e^{-}$,
and the response is linear
up to $\thickapprox80,000$\,$e^{-}$,
with a maximum deviation from linearity of 1\%.
The CCD frame is usually read out at $500$\,kHz;
this takes 8 s, with the data stored
in an 8 MB FITS file.
The readout noise
is $\thickapprox6.5$\,$e^{-}$\,pix$^{-1}$,
with a gain $g\thickapprox2$\,$e^{-}$\,ADU$^{-1}$.
A short (10 s) dark frame exposure has a bias offset
level of $\thickapprox623.4$\,ADU\,pix$^{-1}$,
with rms fluctuations
of $\thickapprox3.6$\,ADU\,pix$^{-1}$,
comparable to the readout noise.
At an operating temperature of -50$^{\circ}$C,
the dark current is
typically $\thickapprox0.030$\,$e^{-}$\,pix$^{-1}$\,s$^{-1}$.
%  
% OLDER TEXT FOR THIS PARAGRAPH FOLLOWS:-
%
%
%
% The pixel well-depth is about $100,000$\,$e^{-}$,
% with a linearity deviation of $\lesssim$1\%.
% The bias level is stable,
% with $\thickapprox3$\,$e^{-}$\,pix$^{-1}$ rms noise.
% The CCD is usually read at $500$\,kHz,
% resulting in $\thickapprox8$\,$e^{-}$\,pix$^{-1}$ rms noise,
% and in a gain $g\thickapprox2$\,$e^{-}$\,ADU$^{-1}$.
%
% Deep image exposures benefit from
% the slower 31 kHz rate,
% giving 2.5 $e^{-}$pix$^{-1}$ rms noise,
% and g$\sim$0.7 $e^{-}$ADU$^{-1}$;
% a sensitivity improvement of 7.8 times for
% weak signal detection, over the 500 kHz rate. 
%
% Each 8 MB FITS file is
% read out in 8 s.
% at 500 kHz.
% The CCD usually operates at -50$^{\circ}$C,
% with a dark current
% of $\thickapprox0.030$\,$e^{-}$\,pix$^{-1}$\,s$^{-1}$.
%
% with a dark current of 
% $\sim$4.8 $e^{-}$pix$^{-1}$
% for a 10 min exposure.
%
%
% START: BOLD type section responding to referee's questions.
% {\bf

\begin{figure}[tb!]
% \setcaptionwidth{3in}
% \includegraphics[scale=0.60]{effarea-345pt.eps}
% \includegraphics[scale=0.42]{andor_QE.pdf}
\includegraphics[scale=0.42]{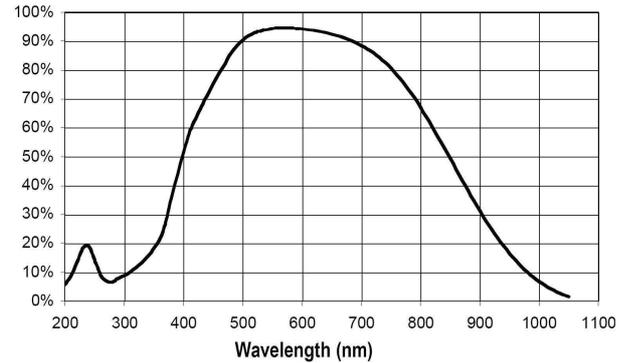}
\caption{ \small
Wavelength sensitivity (QE) of the
iKon-L DW436-BV CCD camera \citep{Andor07}. }
\label{figandorqe}
\end{figure}

The iKon-L sensor is currently used without filters,
and has a nominal peak quantum efficiency (QE) of 95\%
near 5700 \AA \, (in the V-band -- see Figure \ref{figandorqe}).
%
% The effective area of the ZT with the iKon-L camera
% is shown in Figure \ref{figeffarea}.
% This was calculated using the response function (i.e.\ QE)
% of the DW436-BV sensor chip, and also taking into account:
% the 28.7\% areal obscuration from the 54cm diameter secondary
% mirror assembly;
% the wavelength-dependent reflectivity of both aluminised mirrors,
% including a drop of $\sim$ 10\% around $\lambda =$ 8000\AA;
% and a combined 7\% attenuation from the two surfaces of
% each of the corrector-flattener plate, and of the CCD window
% (i.e. a total attenuation of 14\%).
%
The effective area of the ZT with unfiltered Andor CCD
is proportional to their combined spectral response,
which peaks at $A_{\text{EFF}}=3371$\,cm$^{2}$
at $\lambda_{p}=5685$\,\AA,
with a mean wavelength $\lambda_{m}=6225$\,\AA,
and full-width at half-maximum
% (FWHM)
$\Delta\lambda_{\text{FWHM}}=4336$\,\AA;
the overall response is like a `wide-V/R' band,
with relatively low IR and UV sensitivity.

% If the unfiltered ZT system were to image an unextincted
% zero-magnitude A0 V standard star, such as Vega, it would
% detect the `zero-point' flux, which we estimate to be
Assuming Vega's mean spectral irradiance from
$2000$ to $12000$\,\AA\,to be
$S_{0}\thickapprox
2.58\times10^{-9}$\,erg\,cm$^{-2}$\,s$^{-1}$\,\AA$^{-1}$,
we estimate the
zero-point flux\footnote{The zero-point flux detected
by a telescope system is that from an unextincted standard star
of zero magnitude, e.g.\ Vega
% (see \citet{Zombeck90}, \citet{McLean08}).}
(see Zombeck 1990).}
detected by the unfiltered ZT system to be
$F_{0}\thickapprox6.21\times10^{-6}$\,erg\,cm$^{-2}$\,s$^{-1}$.
Integrating over area, the zero-point power detected is
$P_{0}\thickapprox4.95\times10^{-2}$\,erg\,s$^{-1}$,
corresponding to a photoelectron rate of
$\Phi_{0}\thickapprox1.33\times10^{10}$\,s$^{-1}$,
and to a zero-point magnitude
$m_{0}=24.56\pm0.16$;
the effective wavelength of Vega is
$\lambda_{0}\thickapprox5332$\,\AA\, in this system.

Stellar aperture photometry on LIGO trigger wide-field
mosaic frames taken on 2009 July 25 gave an estimated zero-point
of $m_{0}=24.41$ on the USNO-A2.0 magnitude scale,
consistent with our preliminary theoretical estimate.
% }
% END: BOLD face section responding to referee's questions.
%
%
%
\begin{center}
\begin{figure*}[th!] 
% \vspace{-2mm}  \includegraphics[scale=0.35]{Pipeline-2.pdf}
\vspace{-2mm}  \includegraphics[scale=0.35]{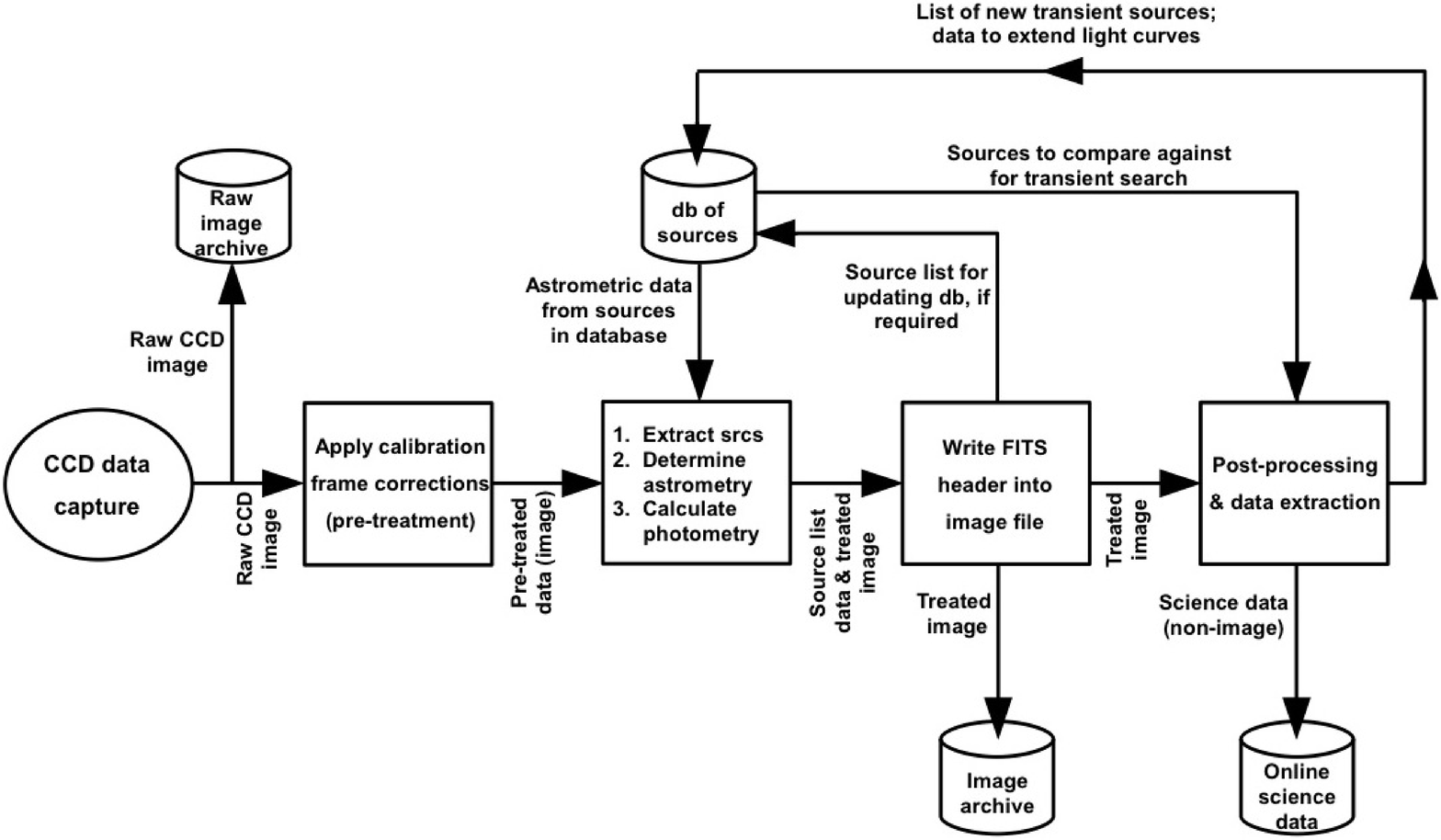}
% \vspace{-2mm}  \includegraphics[scale=0.50]{Pipeline_v05.pdf}
\vspace{-5mm} 
\caption{ \small
Zadko Telescope automated data processing pipeline. } 
\label{figpipeline}
\end{figure*}
\end{center}
\section{Robotic Observatory Control System}
\label{sectrocs}
As part of the TAROT
% \footnote{\url{http://tarot.obs-hp.fr}}
collaboration, the ZT is being robotised so as to be similar
to the existing TAROT robotic telescopes \citep*{Klotz08a}.
The ultimate aim is to make the ZT a node of the TAROT
network, enabling much larger coverage of the transient sky
for participation in several
frontier optical transient science projects.
The centralised system at the core of this network
is a cluster of database servers called
CADOR (Coordination et Analyse des Donn\'{e}es des
Observatoires Robotiques) --- see
\citet*{Bourez-Laas08}, \citet{Klotz08b}. 

Control of the observatory and initial processing of
astronomical images is performed by the
TAROT robotic telescope software system.
This suite of independently running programs comprises
two main components:
AudeLA\footnote{\url{http://www.audela.org/english_audela.php}}
and ROS (Robotic Observatory Software).
There is also 3rd-party software (mostly device drivers) for
the mechanical and ancillary systems of the observatory
(i.e.\ telescope + dome + dome interior + weather station).
For the ZT, the DFM-supplied Telescope Control System (TCS)
is interfaced to ROS via the AStronomy Common Object Model
(ASCOM)\footnote{\url{http://ascom-standards.org} \\
ASCOM is a widely available interface standard for
freeware device drivers commonly used in astronomical
instruments, e.g.\ telescopes, cameras, focusers, and domes.}
protocol.

AudeLA is written mainly in C/C++, and contains
lower-level routines than ROS (which is scripted in Tcl/Tk).
The AudeLA routines can be used by an operator
to directly control the telescope/camera combination,
and to process images.
Alternatively, AudeLA can be driven by
user-created Tcl/Tk scripts, or by other high-level
control software --- such as ROS --- to create
a fully robotic observatory.
The following are the main functions of
the robotic observatory software:

\begin{enumerate} %----LIST

\item Create a prioritised observation schedule for the night.
      Observing requests can be made at any time;
      the queue is dynamically recalculated
      whenever such changes occur.

\item Conduct telescope observations according
      to the queued observing requests and
      priority interrupts (e.g.\ from GRBs).

\item Perform CCD camera operations:
      data capture and transfer to disk.

\item Preprocess raw images.

\item Monitor environmental conditions, and
      activate appropriate software to
      respond to non-benign changes
     (e.g.\ strong wind, rain, high temperature or humidity).

\item Operate Webcam for remote monitoring of observatory.

\item Enable data transfer to/from remote locations.

\item Maintain operational log files.

\end{enumerate} %----END LIST

Many of the above functions rely on AudeLA's
routines to perform the operations
(e.g.\ telescope, dome \& camera control,
image preprocessing --- see Figure \ref{figpipeline}).
After acquiring a raw image from the telescope,
which is stored on local disk, a standard set of processing
operations is then performed on a copy of each image.
Calibration frame corrections
(bias, dark, and flats) are applied.
Cosmic ray artefacts are removed.
Mirror transformations are applied to
put north at the top of the image, and
east to the left.
Stars are identified and catalogued by
SExtractor \citep{Bertin1996}.
Astrometric calibration is then performed
(using, e.g., the USNO-A2.0 catalogue),
followed by photometric determinations.
The image file FITS header is written, and
the resulting processed file is saved to disk.
The CADOR database is also updated with the
new sources measured by SExtractor.
Postprocessing and data extraction are then done.
The observer can specify particular operations
to be applied to the image by the automated pipeline
(e.g.\ checking sources for transient events).
The processed images and any science products
(e.g.\ light curves) are then made available for
the observer to download via the Web.
JPEG images are also produced, for Web page rendering.

The TAROT control system was installed
in 2009 September,
and will be tested in 2010.
This robotic commissioning period will also see
the installation of new environmental monitoring
hardware to ensure that the system is robust.
Also, the dome shutter hardware will be modified
to reduce the risk of mechanical failure,
and to ensure reliable software control. 
We plan to start science programs
in full robotic mode in 2010.

% Data processing pipeline:
% \begin{figure}
% \begin{center}
% \includegraphics[scale=0.45]{Pipeline-2.eps}\vspace{-1cm}
% \caption{ \small
% Automated data processing pipeline. }
% \label{figpipeline}
% \end{center}
% \end{figure}

%=======================================================================
%
\section{Pilot Science Program}
\label{sectpilotsciprog}
In 2009, the telescope and imager performance was tested,
under different observing conditions,
with a pilot science program in optical transient detection.
These first tests served two purposes:
they provided the first sensitivity limits of
the optical and imaging system, and
the first photometric data obtained from
optical transient observations.

%=======================================================================
%
\subsection{ZT Imaging Sensitivity}
A field of faint standard stars centred at
RA(J2000) 16$^{h}$ 11$^{m}$ 04$^{s}$,
Dec.(J2000) +14$^{\circ}$ 56$^{'}$ 36$^{''}$
\citep{Saha05}
was imaged with integration times ranging
from $15$ to $600$\,s,
from 2009 April 20 to May 29.
Stars in the images were detected using
a 2$\sigma$ threshold (i.e.\ 2 $\times$ rms noise).
%
% REMOVED FOLLOWING SENTENCE ABOUT OUR
% (SINGLE, VERY DUSTY) EXTINCTION ESTIMATE.
%
% This field has a maximum elevation
% at the ZT site of $43.5^{\circ}$
% (air mass $X=1.45$);
% the extinction coefficient was estimated
% to be $<$0.44\,mag per airmass.
%
%
% START: BOLD type section responding to referee's questions.
% {\bf
Measured stellar FWHMs have ranged
from 1 to 6 arcsec;
varying from 1 to 3.5 arcsec at zenith,
and 3.5 to 6 arcsec at $45^{\circ}$ elevation.
However, these may tend to overpredict the
atmospheric seeing, due to problems with
collimation and focusing (a good focus cannot
be found over the broad ZT bandwidth).

For these first tests, the limiting magnitude of
the ZT with unfiltered Andor camera was
estimated to be 20.5$\pm$0.1 for integration
times of $300$ to $600$\,s,
under 40\% Moon illumination,
and with average seeing.
With no Moon, the system is sensitive
to sources with $R\thickapprox21$
for a single $180$\,s exposure.
A more typical sensitivity,
for a $2\times180$\,s exposure is $R\thickapprox20$,
for a star $90^{\circ}$ from the Moon
(at 38\% phase; stellar image $\theta_{\text{FWHM}}=4.5$\,arcsec). 
Imaging Target of Opportunity transients showed that
objects with $m\thickapprox23$ are marginally detectable
(SNR $\thickapprox$ 3) using deeper exposures in
favourable observing conditions
(i.e.\ good seeing, dark sky) --- as
demonstrated by the GRB imaging described
in Section \ref{sectgrbs}.

\begin{figure}[th!]
% \setcaptionwidth{3in}
% \begin{center}
% \vbox to100mm{\
% \includegraphics[scale=0.78]{Mick_asteroid_fig1.pdf}
\includegraphics[scale=0.78]{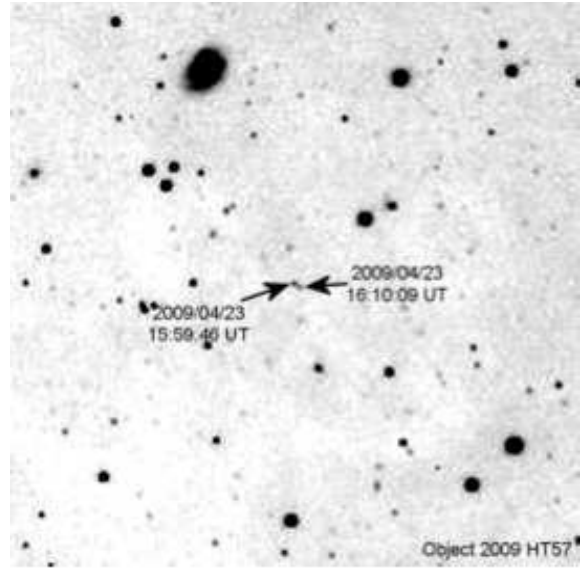}
\caption{ \small
2009 HT57 image, showing the change in position of
this minor planet relative to background stars.
Arrows indicate position at noted times.
Movement is apparent even over a few minutes. }
\label{figmickasteroid}
% \vfil}
% \end{center}
\end{figure}

Our detailed theoretical calculations of the
final SNR --- after careful sky subtraction
and averaging of calibrated CCD frames --- suggest
the following unextincted limiting magnitudes for
unresolved sources detected
(with SNR = 5; image $\theta_{\text{FWHM}}=2$\,arcsec)
by the unfiltered ZT system:
$m\thickapprox24.1$ ($23.8$) for a $10\times100$\,s
stacked integration at
zenith angle $z=0^{\circ}$ ($60^{\circ}$) in New Moon;
$m\thickapprox23.0$ ($22.7$) for a $10\times100$\,s
stacked integration at
zenith angle $z = 0^{\circ}$ ($60^{\circ}$) in Full Moon.
In these estimates, typical values of the atmospheric
extinction coefficient (see Barbieri 2007)
and sky brightness (see McLean 2008) were assumed
in the standard UBVRI filter bandpasses.

On 2009 July 25 CCD frames, a minimum sky brightness
of $m\leq21.16\pm0.04$\,mag\,arcsec$^{-2}$
was measured
(Moon at $5.2^{\circ}$ elevation, $m=-11$,
16\% phase, with scattered clouds present).
%
% Based on this theoretical analysis,
% having no attenuating filter in front of the
% CCD chip should significantly improve the ability
% of the ZT system to detect relatively faint sources.
%
% }
% END: BOLD face section responding to referee's questions.
%
%
%
% \begin{figure}[ht!]
% \setcaptionwidth{3in}
% \includegraphics[scale=0.24]{limiting_mag_plot.eps}
% %\includegraphics[width=7.7cm,height=6cm]{limiting_mag_plot.eps}
% \caption{ \small
% Plot of magnitude limits vs.\ integration time for
% different lunar phases.}
% \label{figmaglimits}
% \end{figure}
%
%=======================================================================
%
\subsection{Minor Planet Searches}
\label{sectminplanets}

In 2009 March, a pilot minor planet search program started
with observations of selected known Near-Earth Objects (NEOs).
Using the Minor Planet Center (MPC) web pages,
for both bright and faint
objects\footnote{\url{www.cfa.harvard.edu/iau/NEO/TheNEOPage.html}},
a shortlist of Near-Earth Asteroids (NEAs),
with magnitudes in the 18--22.5 range,
was compiled for observations in 2009 March and April;
selection was based on considerations of the number of
previous observations,
on information from the MPC urging further observations,
and on the capability of the ZT to detect faint objects.

\begin{table}[tbh!]
\begin{center}
\caption{ \small Some asteroids found using the ZT (2009).}
\label{tabasteroids}
\begin{tabular}{llll}
\hline
2009 ZT     & 2009 ZT    & Previous    & 2010 Feb \\
designation & obs.\ date & obs.\ years & status   \\
\hline
% 2009 FH19$^a$  & Mar 18  & 1991, 1995  & known    \\
2009 FH19$^a$  & Mar 18  &             & arc      \\
2009 FK30$^b$  & Mar 18  & 2004, 2007  & known    \\
2009 FL30      & Mar 21  &             & arc$^e$  \\
2009 GY02      & Apr  2  &             & arc      \\
2009 GZ02      & Apr  2  &             & arc      \\
2009 FN57      & Mar 27  &             & arc      \\
2009 HT57$^c$  & Apr 20  & 2002, 2005  & known    \\
2009 FN59      & Mar 27  &             & lost$^f$ \\
2009 HY81      & Apr 20  &             & arc      \\
2009 HZ81$^d$  & Apr 23  & 1999--2009  & known    \\
2009 JW12      & May 14  &             & lost     \\
2009 KM22      & May 24  &             & lost     \\
\hline
\end{tabular}
\end{center}
% \vspace{-7mm}
\footnotesize{ \noindent
 $^a$ 2009 FH19 is not 30980 = 1995 QU3, as prev. claimed\\
   \hspace*{2mm} by Palomar, and AstDyS database. \\
 $^b$ 2009 FK30 linked by MPC to 2007 UY14 previously \\
 \hspace*{2mm} found in 2004 by Apache Point/SDSS survey. \\
 $^c$ 2009 HT57 linked by MPC to 2005 MH16 previously \\
 \hspace*{2mm} found in 2002 by Palomar/NEAT survey. \\
 $^d$ 2009 HZ81 previously found in 1999 by Palomar. \\
 $^e$ Orbital elements, derived from a limited arc, are \\
 \hspace*{2mm} uncertain --- but predictions are still possible. \\
 $^f$ Orbital elements are too uncertain to make any \\
 \hspace*{2mm} further reliable predictions.  \\
}
% End of \footnotesize
\end{table}

The positions of these selected targets were obtained using
the Minor Planet Ephemeris
Service\footnote{\url{www.cfa.harvard. edu/iau/MPEph/MPEph.html}}
provided by the MPC.
This service also advises which targets require priority
observations, enabling a subset of 4--6 objects to be selected
for each observing session.
Three images of each target were obtained using $300$\,s exposures,
followed by a second set of three images $1$\,hr later.
The IRIS astronomical software
package\footnote{\url{http://www.astrosurf.com/buil/us/iris/iris.htm}}
was used for image preprocessing, astrometric, and
photometric determinations.
The resulting images show the object shifting its position
relative to background stars (see Figure \ref{figmickasteroid}).
These results were subsequently submitted to the MPC to help
refine orbit solutions for use in future searches.

In this program, 12 asteroids were found by chance
(see Table \ref{tabasteroids}).
% at a rate of 0.011 deg$^{-2}$ hr$^{-1}$ of observing.
As of 2010 February, at least 3 of these have
since been
% identified\footnote{\url{http://cfa-www.harvard.edu/cfa/ps/mpc.html}\\
identified\footnote{\url{http://www.minorplanetcenter.org/iau/mpc.html}\\
\hspace*{5.4mm}\url{http://hamilton.dm.unipi.it/astdys/}\\
\hspace*{5.4mm}\url{http://www.hohmanntransfer.com/about.htm}
}
% \footnote{\url{http://hamilton.dm.unipi.it/astdys/}}
% \footnote{\url{http://www.hohmanntransfer.com/about.htm}}
to be recoveries of known asteroids
previously discovered in other programs.
% The remaining 9 may be `new' asteroids (i.e.\ for which
% reliable orbital elements have never previously been determined);
% however, this requires further confirming observations.
Whether any of the remaining 9 asteroids are `new' discoveries
(i.e.\ for which reliable orbital elements
have never been determined) is unknown;
3 of these 9 have since been lost,
and 6 have uncertain elements.
For the 9 objects with known elements,
the semi-major axes range from $1.89$ to $3.17$\,AU,
which suggests that they are main-belt asteroids.

% \begin{figure}
% \setcaptionwidth{3in}
% % \begin{center}
% % \vbox to80mm{\
% \includegraphics[scale=0.5]{Mick_orbit_fig.eps}
% \caption{ \small
%   Orbit of 2009 FH19 3-D representation of the orbit of
%   minor planet 2009 FH19 showing orientation relative to the
%   plane of the Solar System. Labels indicate positions of
%   Solar System bodies at April 1, 2009. }
% %\label{figmickorbit}
% %\vfil}
% %\end{center}
% \end{figure}

%=======================================================================
%
\subsection{Optical Transient Searches}
\label{sectgrbs}

TAROT employs two fully robotic 0.25-m telescope systems
located in Calern (France) and La Silla (Chile),
mainly for the rapid follow-up of GRBs triggered by satellites.
GRBs are high-energy transients, lasting
from milliseconds to hundreds of seconds.
Observations of these emissions provides a probe of the
extreme physics associated with the largest explosions
in the Universe.
Optical flares have been observed by TAROT and other
telescopes, in some cases tens of seconds
after the trigger, allowing new insight into the
prompt emission regime of GRBs \citep{Klotz09}. 

\begin{figure}[ht!]
%\setcaptionwidth{3in}
% \includegraphics[scale=0.45]{hist_decs1.pdf}
\includegraphics[scale=0.45]{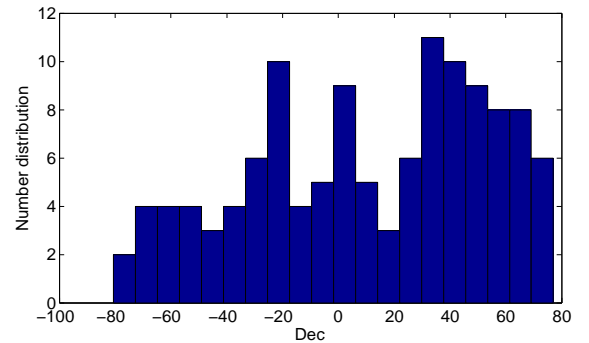}
\caption{\small Observed GRB optical afterglow distribution
in declination. }
\label{figgrbhist}
% It is clear there is a deficit of afterglow observations
% at negative declinations.
% The main cause of this reduction is a selection bias from
% the much smaller number of telescopes engaged in
% GRB optical follow-up at southern latitudes. }
\end{figure}

The record of `missed' GRB optical afterglow observations
provides insight into the impact that the ZT will have on
obtaining photometry on bursts at negative declinations.
In the following, we show how the historical record of
satellite-triggered optical transient observations has
revealed a strong bias in the number of GRB afterglows
observed at different declinations.

About half of the observations triggered and localised by
the {\it Swift} satellite have not been observed with any
optical emission, despite deep follow-up observations.
To fully capitalise on the positional information
from {\it Swift}, and other satellite detectors, requires
follow-up by optical telescopes of as many triggers as possible.
As the continental land mass is strongly biased to the northern
hemisphere, most rapid-response telescopes engaged in
optical transient science are located in Europe, Asia, and
North America. This has resulted in a deficit of GRB
optical afterglow (OA) science at declinations south
of $-60^{\circ}$.
Figure \ref{figgrbhist} highlights the drop in GRB OA follow-up
in the southern hemisphere, by plotting the distribution of
GRB OA discoveries with declination.
Without OA detections, redshifts cannot be secured
to these bursts.
\citet{Coward08} showed how such selection effects can modify
the {\it observed} spatial distribution of GRB optical properties.

It is clear from this distribution that a niche role exists
for metre-class telescopes in Australia, including the ZT,
to help fill the GRB OA (and other transient) deficit at
negative declinations.  

After installing the Andor camera on the ZT in 2008 September,
a Target of Opportunity (ToO) pilot program was started,
aimed at GRB OAs. Even though the facility was not enabled for
robotic response when this program
started\footnote{In 2009 September, the French ROS system was
installed, enabling rapid response to GRB satellite triggers;
this system will be tested over the following year.},
the communication infrastructure was in place
to receive external triggers,
and to manually intervene in any scheduled observing.
Just weeks after installing the camera,
the ZT was imaging {\it Swift} satellite-triggered
GRB optical afterglows reported in online
circulars of the GRB Coordinates Network
(GCN)\footnote{\url{http://gcn.gsfc.nasa.gov}}.
Below, we summarise these pilot observations
(see Table \ref{tabgrbs}):

\begin{table}[thb!]
\begin{center}
\caption{ \small
ZT GRB ToO observations (2008--2009).}
\label{tabgrbs}
% \caption{ \small
%    GRB Target of Opportunity observations
%    (2008 Nov--2009 Nov 2009)
%    used for testing the communications, telescope pointing,
%    camera and control system. }
\begin{tabular}{llll}
\hline
 GRB Number    & ZT estim.\                  & $z$   & ZT Obs.\ \\
               & magnitude                   &       & Report   \\
\hline
GRB 081118$^a$ & 20.9$\pm$0.5                & 2.58  & GCN 8675 \\
GRB 090205     & 23.2$\pm$0.3                & 4.65  & GCN 8976 \\
GRB 090313     & 18.0$\rightarrow$18.8       & 3.38  & GCN 8996 \\
% GRB 090419     & $>21$                       &       &          \\
GRB 090509     & $>17.6^{c}$                 &       & GCN 9363 \\
GRB 090516     & 18.9$\rightarrow$20.3       & 4.11  & GCN 9380 \\
GRB 090927$^b$ & 19.8$\rightarrow$20.0$^{c}$ & 1.37  & GCN 9956 \\
GRB 091127$^d$ & 18.9$\pm$0.1                & 0.49  & GCN 10238 \\
\hline
\end{tabular}
\end{center}
% \vspace{-7mm}
\footnotesize{ \noindent
 $^a$ GRB 081118 is one of the most distant transients imaged \\
   \hspace{4mm} with a video camera (the Adirondack Stellacam). \\
 $^b$ Possibly a short GRB. \\
 $^c$ Based on R-band magnitudes of USNO catalogue stars. \\
 $^d$ GRB with unusually slow-decaying OA.
}
% End of \footnotesize
\end{table}

\begin{figure}[hbt!]
% \setcaptionwidth{3in}
% \includegraphics[scale=0.45]{g090516n-28x50s-004.eps} 
% \includegraphics[scale=0.42]{grb090927_zadko1.pdf}
\includegraphics[scale=0.42]{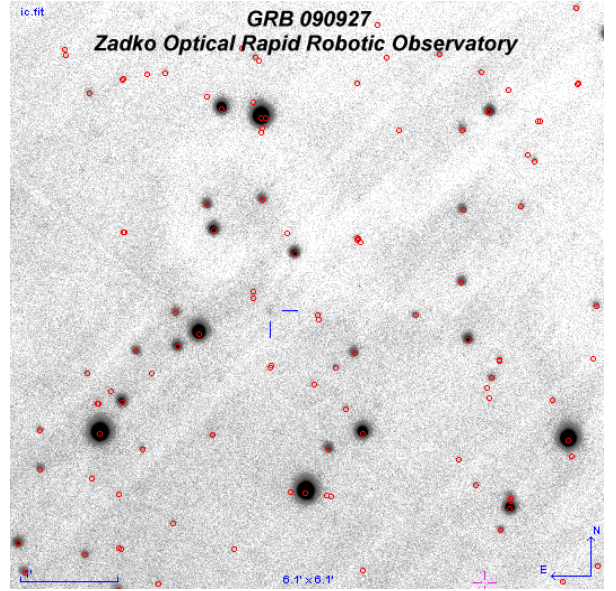}
\caption{ \small GRB 090927 optical afterglow, shown by the pointer
marks near the image centre.
Red circles are known stars from the NOMAD1 catalog.
This ZT image is a sum of $2\times180$\,s exposures
taken $2.6$\,hr post-{\it Swift} trigger.
Stellar images have $\theta_{\text{FWHM}}=1.8$\,arcsec.
The afterglow at this time had $R=20.0 \pm 0.5$
(image processed by TAROT collaboration). }
\label{figgrbzorro}
% \caption{Image of GRB090516 (arrow) obtained by stacking
% $28\times50$ s exposures. }
% \label{figgrbstack}
\end{figure}

\begin{itemize}
\item {\bf GRB 081118:}
This was the first optical transient imaged by the ZT.
% The acquisition of the images was rather dramatic
% in that CCD camera control computer failed
% after the imaging commenced.
Imaging started $1.96$\,hr
post-{\it Swift} burst trigger, using
an unfiltered, integrating CCD video camera,
% (the Stellacam),
and continued for $1$\,hr.
The OA candidate reported in GCN 8529 was seen
in a selected stack of $40\times5$\,s exposures,
though it was near the detection limit.
Using nearby USNO-B1 stars, the OA magnitude was
found to be $20.9 \pm 0.5$ (GCN 8675).
Confirmations of this OA candidate were made by
the 2.2-m GROND telescope and the VLT;
a VLT spectroscopic redshift of $z = 2.58$ was
determined $8$\,hr after the ZT observation.

\item {\bf GRB 090205:}
This was a late-time observation;
imaging started $17.6$\,hr
post-{\it Swift} trigger,
with the unfiltered Andor camera,
and continued for $4460$\,s.
In a $1100$\,s stacked image, a candidate OA
with $23.2 \pm 0.3$ white magnitude was
marginally detected (SNR = 3.3) at the
enhanced XRT position reported in GCN 8885.
This OA magnitude was estimated using
nearby USNO-B1 stars (GCN 8976).
As the source was near the detection limit,
the same field was imaged 2 weeks later:
the candidate OA was not detected,
which is consistent with a fading GRB as
a possible cause of this transient.

\item {\bf GRB 090313:}
Imaging of the field reported in GCN 8980
started $4.26$\,hr post-{\it Swift} trigger,
with significant sky brightness from the Moon.
The OA decayed from $18.02 \pm 0.08$
to $18.84 \pm 0.12$ in $8500$\,s (GCN 8996).

% \item {\bf GRB 090419:}
% An upper limit of 21 mag was estimated
% for this OA (not reported).

\item {\bf GRB 090509:}
Imaging of the field reported in GCN 9325
started $8$\,hr post-{\it Swift} trigger;
$48\times5$\,s exposures were
made over $14$\,min.
Due to sky brightness
from a full Moon,
no candidate OA was found at
the GROND position (GCN 9326), nor at
the BAT-refined position (GCN 9349),
with an upper limit of $R = 17.6$
estimated from nearby USNO-B1 stars (GCN 9363).

\item {\bf GRB 090516:}
Imaging started $275$\,min post-{\it Swift} trigger,
and continued for $131$\,min.
An OA candidate was found within the XRT error circle,
and it faded from $18.9 \pm 0.5$ to $20.26 \pm 0.14$
over $21$\,min (GCN 9380).

\item {\bf GRB 090927 -- a possible short GRB:}
Imaging of the field reported
by {\it Swift}/UVOT (GCN 9946)
started $110$\,min post-{\it Swift} trigger, and
continued for $50$\,min in piloted robotic mode.
A fading source was found within the XRT error
circle -- see Figure \ref{figgrbzorro}.
Preliminary photometry of this OA candidate
gave R-band magnitudes of:
$19.8 \pm 0.5$ ($540$\,s exposure) and
$20.0 \pm 0.5$ ($360$\,s),
at post-burst times
of $126$ and $158$\,min,
respectively (GCN 9956).

\item {\bf GRB 091127 -- with a slowly-decaying OA:}
This was a bright $15.0$\,mag source which triggered
after dawn, but had an unusually slow decay time.
Using a Pentax K200D digital SLR camera
mounted at the Cassegrain focus, late-time imaging
of the field reported in GCN 10191
started $18.5$\,hr post-{\it Swift} trigger,
and continued for $28$\,min; during this time,
the Moon set, and weather conditions were good.
A faint source was found at the OA candidate location
reported in GCN 10199. Photometry on a co-added sum
of 4 CCD images (of $806$\,s total exposure time)
gave an OA magnitude of $18.9 \pm 0.1$
(GCN 10238).

\end{itemize}

%=======================================================================
%
\section{Future Development}
\label{sectfuturedev}

%=======================================================================
%
\subsection{Instrument Upgrade}
\label{sectinstup}
The Guide Acquire Module (GAM) supplied with the DFM telescope
places the optimal image plane too close to the current optical
mounting plate to allow a large filter wheel to be placed in
front of the current Andor camera.
There are several solutions to this problem,
depending on available funds;
% a desirable one would be to replace the GAM with
a particularly attractive one would be to replace the GAM
with a new optical mounting structure, to ensure that there
is space for mounting the filter wheel, electronic focuser,
and Andor CCD camera.
The new design will allow for three optical paths,
for:
filtered CCD imaging,
an integrating video camera,
and a spectrograph. 

Subject to funding, a new CCD camera, filter wheel,
and filter system will be installed.
One of the imagers being considered is
the Fairchild Peregrine 486 CCD camera,
which can be matched to the image focal plane
of the existing optical mount.
It has a $61.44\times61.44$\,mm chip,
with an imaging FOV of $1^{\circ}$,
or 7 times the area of the existing camera;
by significantly increasing the survey speed,
this camera would extend the science capabilities
of the ZT in the field of transient searches.

%=======================================================================
%
\subsection{ZT Science Linkages}
\label{sectztlinks}

ZT research is tightly linked to
several existing and upcoming
international programs in transient astronomy.
The ZT team is part of the TAROT collaboration,
which already uses 2 other very rapid response
telescopes (in France and Chile).
We are working closely with the GW astronomy
community (the LIGO Scientific Collaboration)
to use the ZT for optical follow-up of
candidate GW events, and to survey galaxies
within the LIGO sensitivity horizon for coincident
optical-GW transients.
Our TAROT collaboration is also working with the
the Antares deep-sea neutrino telescope to follow up
on possible neutrino events.
These projects are at the frontier of
multi-messenger astronomy, and
it is very timely for the ZT to play a role
when these projects are just starting.

The U.S. GW observatory LIGO started searching for
GW transient signals, with unprecedented sensitivity,
in late 2009 \citep{Abbott07}.
Despite the detection of significant numbers of
candidate GW transient events in LIGO data
in recent science runs, the SNR of these correlated
signals is relatively small.
It is difficult to confirm that these events are
astrophysical in origin, rather than environmental \citep{Fox05}.
However, the association of such an event with an
astrophysical transient observed in the optical
would dramatically improve the confidence of GW detection.
The ZT, with our partners in the TAROT collaboration and
other transient telescopes, will be testing and developing
procedures to search for coincident GW and optical transients.  

The Australian SKA Pathfinder (ASKAP),
currently under construction, is a revolutionary new design
of radio telescope, capable of monitoring $30$\,deg$^{2}$ of sky
instantaneously, forming images with $10$\,arcsec resolution
\citep{Johnston08}.
ASKAP will have both wide sky coverage and high sensitivity,
plus the ability to locate transient radio sources to
high precision. However, for fast radio transients
that appear and disappear on timescales as short as a millisecond,
searches will initially be incoherent (i.e.\ uncorrelated);
this will only provide low-resolution positional information,
although offline postprocessing of the raw data can also provide
high-resolution positions. Incoherent detection will provide
ASKAP triggers with positional accuracy of $1^{\circ}$,
which is well-matched to the capability of the ZT,
% assuming a modest camera upgrade to achieve a 1$^{\circ}$ FOV.
when it is upgraded to achieve a $1^{\circ}$ FOV.
The proximity of the ZT to ASKAP also provides a unique advantage
in the follow-up of short-lived ASKAP radio transients.

%=======================================================================
%
\subsection{Remote Access Astronomy in Education}
\label{sectremasteduc}
The ZT has initiated a novel public outreach program that
allows small groups to obtain CCD images on site
for a few hours every month.
This program is part of a broader science outreach project.
Furthermore, projects are being developed that will allow students
to participate in ZT science, and to design and implement projects
which will allow them to make useful scientific contributions.
The project list is long, but includes:
% \newline

\begin{itemize}
\item Minor planet and NEO recovery, and discovery.
\item Supernova surveys, discovery, and follow-up.
\item GRB and other transient event follow-up.
\end{itemize}

As the global awareness of the importance of astronomy
as a context for teaching science continues to increase,
educators are now able to re-engage in observational astronomy
through the use of remote access to robotic telescopes.
In parallel with this trend, wide-field instruments are
generating a wealth of data directly available to the public
for analysis. There are very practical reasons for this,
namely that the huge data volumes make it difficult
to search for faint optical transients.
This represents an opportunity for science educators
to tap into at relatively small cost, with the benefit of
contributing to real science. 

The ZT is primed to take advantage of these opportunities
in education, training, and outreach.
The ZT observatory control system, using the TAROT robotic system,
allows science data to be acquired remotely using a web-based interface.
Our French collaborators have successfully used this system
for education and outreach.
The ZT will act as a node in a distributed network of
robotic telescopes, allowing the opportunity for students
in Europe to acquire image data from the ZT.
Conversely, local students can acquire data from the other
networked telescopes in France and Chile. 

An education research project is being developed, as part of
an ARC Linkage grant, aimed at measuring the effectiveness
of student learning via remote access astronomy;
with astronomy as the context, this will provide valuable data
on how different modes of learning science influence
students' choices to continue science education from
high school to tertiary level.
%
%=======================================================================
%
%
\section*{Acknowledgements} %If needed
D.~M. Coward is supported by ARC grants DP 0877550, LP 0667494,
and by the University of Western Australia (UWA).
The ZT facility is managed by the School of Physics, UWA,
the International Centre for Radio Astronomy Research (ICRAR),
and the Australian International Gravitational Research Centre (AIGRC).
We also acknowledge the Centre for Learning Technology (CLT) at UWA,
and the Western Australian Department of Education and Training (DET),
for supporting a science training and outreach program.
%

%\end{multicols}


\begin{thebibliography}{}
% References are listed as in the following example,
% for more examples, please see the PASA Style Guide
% \bibitem[Smith, Jones, \& Brown(Year)Smith et al.]{example}
%  Smith, A.~B., Jones, C.~D., Brown, E.~F. Year, Journal, Volume, Page
%
% (Journal articles)
%
\bibitem[Abadie et al.(2010)]{Abadie10}
Abadie, J., et al. (LIGO Scientific Collaboration and
the VIRGO Collaboration) 2010,
LSC Document LIGO-P0900125,
e-print (arxiv:1003.2480v2 [astro-ph.HE])
% ``Predictions for the Rates of Compact Binary Coalescences Observable
%   by Ground-based Gravitational-wave Detectors''
%
\bibitem[Abbott et al.(2007)]{Abbott07}
Abbott, B., et al. 2007, PhRvD, 76, 062003
% (Physical Review D)
% ``Search for gravitational wave radiation associated with the pulsating tail
%   of the SGR 1806-20 hyperflare of 27 December 2004 using LIGO''
%
\bibitem[Andor Technology(2007)]{Andor07}
Andor Technology PLC, 2007. iKon-L DW436-BV Technical Document
(Belfast, N.\ Ireland: Andor Technology PLC)
% (Andor Technology 2007, iKon-L DW436-BV CCD Camera)
% (Instrument documentation)
%
% ASCOM reference needed.
% URL from Alan Imerito:
%   http://ascom-standards.org
%  (this URL is supplied in a footnote).
%
\bibitem[Barbieri(2007)]{Barbieri07}
Barbieri, C. 2007, Fundamentals of Astronomy
(Boca Raton FL, USA: CRC Press, Taylor \& Francis), 292
%
% Typical atmospheric extinction coefficients at various
% optical wavelengths (for Mauna Kea).
% These need to be scaled by the scale lengths of
% the overlying atmospheres, to give the corresponding
% coefficients for ground level observing.
%
\bibitem[Bertin \& Arnouts(1996)Bertin \& Arnouts]{Bertin1996}
Bertin, E., \& Arnouts, S. 1996, A\&AS, 117, 393
% (Astronomy & Astrophysics Supplement)
% ``SExtractor: Software for source extraction''
%
\bibitem[Bourez-Laas, Vachier, Klotz, Damerdji, Bo\"{e}r(2008)
Bourez-Laas et al.]{Bourez-Laas08}
Bourez-Laas, M., Vachier, F., Klotz, A., Damerdji, Y., \& Bo\"{e}r, M. 2008,
Proc. SPIE, 7019, 701918
% (Proceedings of the SPIE --
%  Society of Photographic Instrumentation Engineers, founded 1955)
%  Advanced Software and Control for Astronomy II.
%  Edited by Bridger, Alan; Radziwill, Nicole M.
%  Proceedings of the SPIE, Volume 7019, pp. 701918-701918-9 (2008).
% ``CADOR and TAROT: a virtual observatory''
%
\bibitem[Coward et al.(2008)]{Coward08}
Coward, D.~M., Guetta, D., Burman, R.~R., \& Imerito, A. 2008, MNRAS, 386, 111
% (Monthly Notices of the Royal Astronomical Society)
% ``Where are the missing gamma-ray burst redshifts?''
%
% \bibitem[Dodson(2009)]{Dodson09}
% Dodson, R. 2009, (private communication)
% (Need a good, recent reference/URL to ASKAP).
%
% CSIRO ATNF ASKAP Website
% http://www.atnf.csiro.au/SKA/
%
%
\bibitem[Fox et al.(2005)]{Fox05}
Fox, D.~B., et al. 2005, Nature, 437, 845
% (Nature journal)
% ``The afterglow of GRB 050709 and the nature of the
%   short-hard gamma-ray bursts''
%
% \bibitem[Green (1992)]{Green92}
% Green, D.~W.~E. 1992, International Comet Quarterly, 14, 55
% (International Comet Quarterly)
% ``Magnitude Corrections for Atmospheric Extinction''
%
% \bibitem[Horne(2007)]{Horne07}
% Horne, J. 2007, S\&T, 114(3), 64
% (Instrument magazine article)
% (Sky & Telescope magazine; Sky Publishing Corp., USA)
% ``Sky & Telescope, 2007, VOL 114; NUMB 3, pages 64-67''
% ``S&T Test Report Next-Generation Video: Adirondack's StellaCam3''
%
% http://www.astrovid.com/prod_details.php?pid=3306
% http://direct.bl.uk/ (British Library Direct)
%
%
% \bibitem[Jablonski(1992)]{Jablonski92}
% Jablonski, M. 2003, euphOrbit
% (Software)
%
%
% \bibitem[Johnston et al.(2007)]{Johnston07}
% Johnston, S., et al. 2007, PASA, 24, 174
% Publications of the Astronomical Society of Australia,
% Volume 24, Issue 4, pp. 174-188
% ``Science with the Australian Square Kilometre Array Pathfinder''
%
\bibitem[Johnston et al.(2008)]{Johnston08}
Johnston, S., et al. 2008, ExA, 22, 151
% Experimental Astronomy, Volume 22, Issue 3, pp.151-273
% ``Science with ASKAP. The Australian square-kilometre-array pathfinder''
% ASKAP 1.4 GHz 12-m single-dish FOV = 30 deg^2 (with a 30-beam FPA).
% Each ASKAP single dish beam pixel will have a FOV = 1 deg^2.
%
% \bibitem[Keller(2007)]{Keller07}
% Keller, S.~C., et al. 2007, PASA, 24, 1
% (Publications of the Astronomical Society of Australia;
%  continuation of PASAu)
% ``The SkyMapper Telescope and The Southern Sky Survey''
%
\bibitem[Klotz, Bo\"{e}r, Eysseric, Damerdji, Laas-Bourez, Pollas,
Vachier(2008a)Klotz et al.]{Klotz08a}
Klotz, A., Bo\"{e}r, M., Eysseric, J., Damerdji, Y., Laas-Bourez, M.,
Pollas C., \& Vachier F. 2008a, PASP, 120, 1298
% (Publications of the Astronomical Society of the Pacific)
% ``Robotic Observations of the Sky with TAROT: 2004-2007''
%
\bibitem[Klotz, Vachier, \& Bo\"{e}r(2008b)Klotz et al.]{Klotz08b}
Klotz, A., Vachier, F., \& Bo\"{e}r, M. 2008b, AN, 329, 275
% (Astronomische Nachrichten)
% ``TAROT: Robotic observatories for gamma-ray bursts and other sources''
%
\bibitem[Klotz et al.(2009)]{Klotz09}
Klotz, A., Gendre, B., Atteia, J.~L., Bo\"{e}r, M., Coward, D.~M.,
Imerito, A.~C. 2009, ApJ, 697, L18
% (Astrophysical Journal Letters)
% ``Observation of Correlated Optical and Gamma Emissions from GRB 081126''
%
% \bibitem[Klotz, Bo\"{e}r, Atteia, \& Gendre(2009)Klotz et al.]{Klotz09b}
% Klotz, A., Bo\"{e}r, M., Atteia, J.L., \& Gendre, B. 2009, AJ, 137, 4100
% (Astronomical Journal)
% ``Early Optical Observations of Gamma-Ray Bursts
%   by the TAROT Telescopes: Period 2001-2008''
% 
% \bibitem[Law et al.(2009)]{Law09}
% Law, N.~M., et al. 2009, PASP, 121, 1395
% (Publications of the Astronomical Society of the Pacific,
%  Volume 121, issue 886, pp.1395-1408)
% ``The Palomar Transient Factory: System Overview, Performance,
%   and First Results''
% Law, N.~M., et al. 2009, (arXiv:0906.5350)
%
\bibitem[McLean(2008)]{McLean08}
McLean, I.~S. 2008, Electronic Imaging in Astronomy:
Detectors and Instrumentation (2nd ed; Chichester, UK:
Springer-Praxis Publ.\ Ltd.), 46
%
% Vega spectral irradiances in standard Kron-Cousins filter
% bands (p. 347) - not used in versions since 9e of this paper.
%
% Typical sky brightnesses in mag/arcsec^2 in UBVRI bands,
% for both New Moon and Full Moon conditions (p. 46).
%
%
% \bibitem[Rau et al.(2009)]{Rau09}
% Rau, A., et al. 2009, PASP, 121, 1334
% (Publications of the Astronomical Society of the Pacific,
%  Volume 121, issue 886, pp.1334-1351)
% ``Exploring the Optical Transient Sky with the Palomar Transient Factory''
% Rau, A., et al. 2009, (arXiv:0906.5355)
%
\bibitem[Saha et al.(2005)]{Saha05}
Saha, A., Dolphin, A.~E., Thim, F., \& Whitmore, B. 2005, PASP, 117, 37
% (Publications of the Astronomical Society of the Pacific)
% ``Faint BVRI Photometric Sequences in Selected Fields''
%
\bibitem[Sathyaprakash \& Schutz(2009)]{Sathyaprakash09}
Sathyaprakash, B.~S., \& Schutz, B.~F. 2009,
Living Reviews in Relativity, 12(2), 99-102, (arxiv:0903.0338v1 [gr-qc])
% ``Physics, Astrophysics and Cosmology with Gravitational Waves''
%
\bibitem[Zombeck(1990)]{Zombeck90}
Zombeck, M.~V. 1990, Handbook of Space Astronomy \& Astrophysics
(2nd ed; Cambridge, UK: Cambridge Univ.\ Press), 100
%
% Absolute spectral irradiances for Vega in UBVRI Johnson-Morgan filters.
%
%
\end{thebibliography}
\end{document}